\documentclass[prl,aps,floatfix, twocolumn,superscriptaddress,showpacs]{revtex4} 
\usepackage{amsmath,bm,graphicx}
\usepackage{amssymb}
\usepackage{amsfonts}
\usepackage{color}
\begin{document}

\title{Power-law distributions of particle concentration in free-surface flows}

\author{Jason Larkin}
\email[Corresponding Author : ]{jml37+@pitt.edu }
\affiliation{Department of Physics \& Astronomy, University of Pittsburgh, Pittsburgh, PA 15260, USA.}

\author{M. M. Bandi}
\affiliation{Center for Nonlinear Studies and Condensed Matter \& Thermal Physics Group, Los Alamos National Laboratory, Los Alamos, NM 87545, USA.}

\author{Alain Pumir}
\affiliation{Laboratoire de Physique de l'Ecole Normale Sup{\'e}rieure de Lyon, Lyon 69364, France.}

\author{Walter I. Goldburg}
\affiliation{Department of Physics \& Astronomy, University of Pittsburgh, Pittsburgh, PA 15260, USA.}

\date{\today}

\begin{abstract}
Particles floating on the surface of a turbulent incompressible fluid accumulate along string-like structures, while leaving large regions of the flow domain empty. This is reflected experimentally by a very peaked probability
distribution function of $c_r$, the coarse-grained particle concentration at scale $r$, around $c_r = 0$, with a power-law decay over two decades of $c_r$, $\Pi (c_r) \propto c_r^{-\beta_r}$. The positive exponent $\beta_r$
decreases with scale in the inertial range, and stays approximately constant in the dissipative range, thus indicating a qualitative difference between the dissipative and the inertial ranges of scales, also visible in the first moment of $c_r$.
\end{abstract}


\maketitle

\section{1. Introduction}

Consider the introduction of passive, neutrally buoyant tracer particles into an incompressible turbulent flow, for example ink injected into a vigorously stirred tank of water. After  a transient mixing period, the dye becomes uniformly distributed.  Further  stirring will have no obvious effect; the dye  particles will be rearranged on a microscopic scale, but  on larger scales, the concentration distribution will remain uniform.

\begin{figure}[h]
$\begin{array}{c}
\includegraphics[width = 3.5 in]{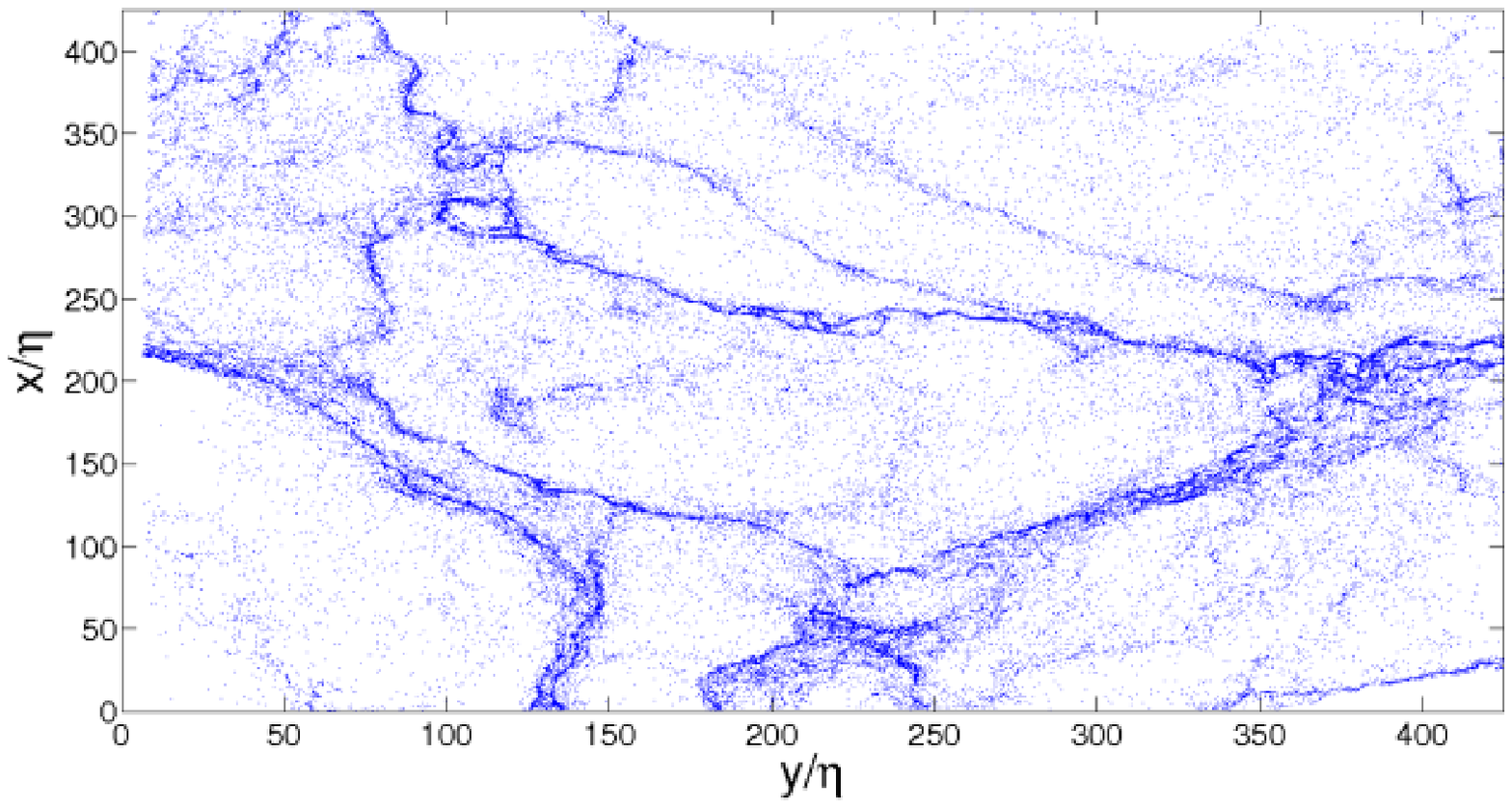} \\
\includegraphics[width = 3.5 in]{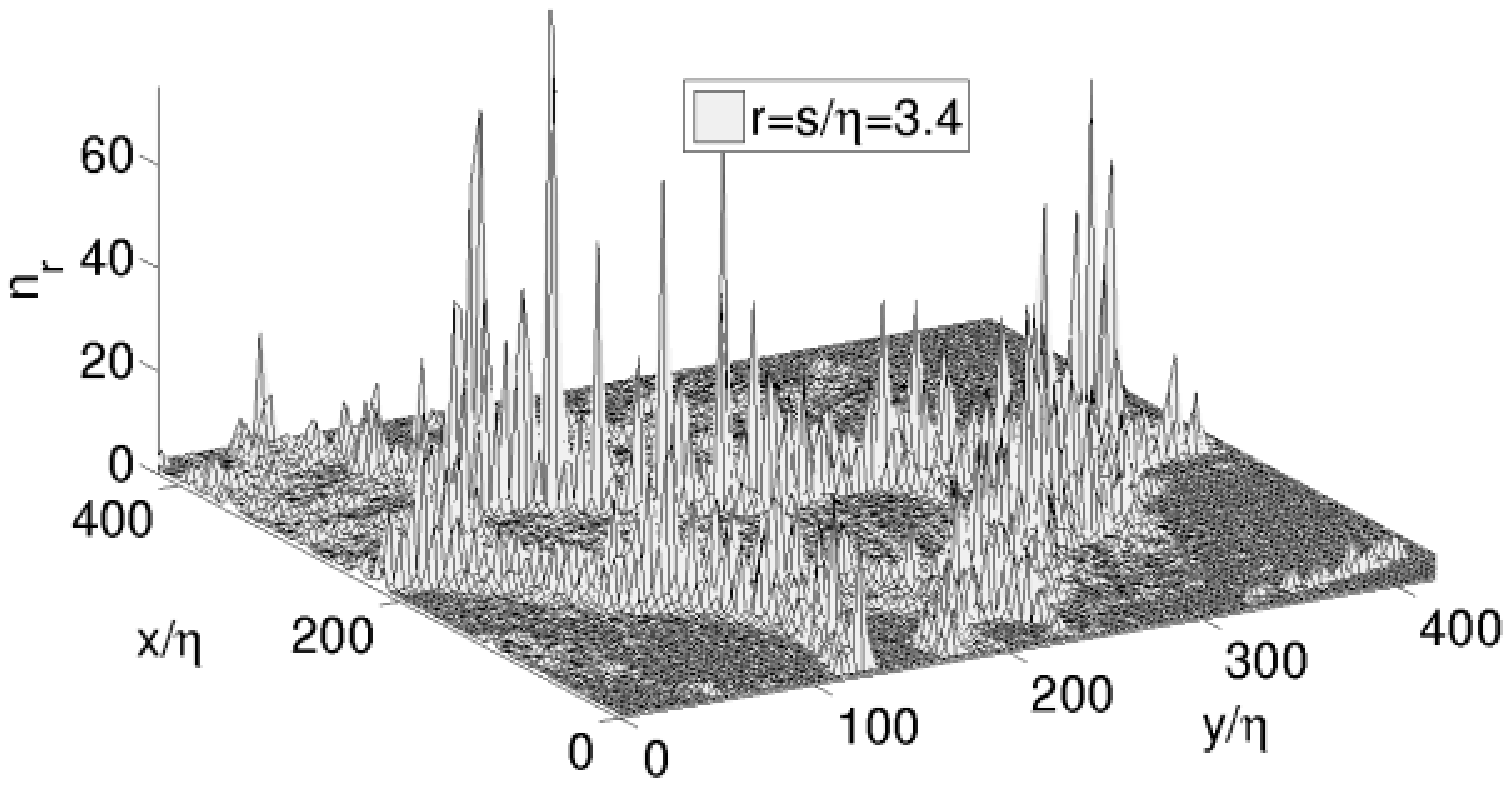}
\end{array}$
\caption{(Color online) Top panel: Visualization of the steady-state particle distribution shows string-like coagulations surrounded by (relatively) depleted regions. Bottom panel: The Eulerian concentration field $n_r$ in the steady-state shows a very intermittent distribution, with very large spikes concentrated along string-like structures, surrounded by vast depleted regions which are nearly flat. The $x$ and $y$ axis are in units of $\eta$ (see text). The coarse-graining scale chosen here is $r = 3.4$. Decreasing the coarse-graining scale $r$ increases the intermittent nature of the particle concentration. }
\label{particles}
\end{figure}

Now consider a different experiment in which the contaminant
particles are lighter than the host fluid, so that they float on the fluid-air
interface. This situation, which occurs naturally for certain classes of
phytoplankton or contaminants \cite{R1926,S1949,A2000}, has been studied
only recently in the laboratory \cite{CDGS2004}.
Even when the floaters are initially distributed
uniformly on the surface of the  host fluid, which  is continuously  stirred
before and after the surface is covered, the  concentration
distribution of the floaters will not remain uniform.  Rather, the floaters
will cluster into string-like structures, see Fig.\ref{particles}.
The underlying turbulence, which drives the motion of the floaters, will
cause the string-like distribution to
fluctuate in  space and time, but there will be no tendency toward
concentration uniformity.

This clustering is not caused by an interaction between the
floaters, nor is it produced by surface tension effects or by wave motion.
It is seen in simulations of the turbulent motion of the underlying fluid
when these complicating effects are not included. In the simulations
\cite{BDES2004,CDGS2004}, the fluid, assumed to be governed by the
incompressible Navier-Stokes equation, is driven at large spatial scales
to turbulence. The floaters merely follow the fluid motion and sample the
horizontal velocity components at the surface; they are therefore treated as
inertia-free.


Here we study the Lagrangian evolution of the steady-state tracer concentration of the floaters at the interface between air and water. Although the underlying turbulence is incompressible, the particles which are lighter
than water are constrained to move only along the two-dimensional interface, they cannot follow water into the bulk. The tracers 
flee fluid up-wellings (sources) and cluster along the fluid down-wellings (sinks) and form a compressible system.
From a dynamical systems perspective the effect seen in Fig. \ref{particles}
results from a competition between stretching along one spatial direction, and
contraction over another one. The particle motion is characterized by a positive
and a negative Lyapunov exponents, with the sum being negative \cite{BGH2004,DP2008}.

The effect of the surface flow that induces floater clustering is quantified by the two-dimensional compressibility: ${\cal C}=\left\langle\left(\stackrel{\rightharpoonup}{\nabla }_{2} \cdot \stackrel{\rightharpoonup}{v}\right)^{2} \right\rangle / \left\langle \left(\stackrel{\rightharpoonup}{\nabla }_{2} \stackrel{\rightharpoonup}{v}\right)^{2} \right\rangle $. For isotropic turbulence, ${\cal C}$ lies between $0$ (incompressible fluids) and unity (potential flows). Past experiments and Direct Numerical Simulations (DNS) have measured ${\cal C} \approx 1/2$ \cite{CDGS2004}, making this a strongly compressible system. The accumulation of particles in surface flows is reminiscent of the physics leading to clustering of inertial particles in incompressible flows \cite{BDL2006, FP2004}. There the clustering arises due to inertia of advected particles whose trajectories deviate away from the stream lines. In the present case of free-surface turbulence, the floaters are passive and closely follow the local flow field, the clustering arises purely because their motion is constrained to the air-water interface. The fluid flow acts as a dynamical system on the floating particles \cite{BDES2004,BGH2004,BDL2006,PFL2006, BGC2006,BCG2008}. At scales below the dissipative scale of the turbulence, $\eta$, the flow is viscous (smooth), and theories based on the stretching of fluid elements lead to the prediction that the particle distribution is multifractal \cite{BGH2004,PFL2006}. For particle distributions in the inertial range where the velocity field is not smooth, the only available study comes from a surface flow model that suggests a multi-fractal particle distribution for both the inertial and dissipative ranges \cite{DP2008}.

We study the concentration statistics of particle clusters in free-surface flows as a function of the dimensionless coarse-graining scale $r$ in the steady state. The distribution of the coarse-grained concentration $n_r$ over a domain of scale $r$, is very intermittent (Fig. \ref{particles}), the more so as the scale $r$ decreases. This can be seen in Fig. \ref{particles}, where particles have been expelled from large regions of the flow by surface sources and have accumulated along dense line-like surface sinks. Intermittency can be quantified by measuring the
moments of $n_r$ averaged over all points in space (Eulerian representation). In the present work, we instead investigate the Lagrangian concentration $c_r$ averaged over a domain of size $\sim r$ around {\it each } particle, which is equivalent to the Eulerian quantity (see Eq. \ref{Eul_Lag}).

The first main result of this work is the observation of two different scale-free regions for the first moment of $c_r$, corresponding to the dissipative and inertial ranges, in qualitative agreement with \cite{DP2008}. The PDF of
the coarse-grained concentration $\Pi(c_r)$ exhibits a steep (discussed below) power-law behavior at small values of $c_r$, characterizing the regions with a very low particle density. PDFs with power-law behavior are frequently encountered in nature in many different contexts \cite{N2005}, usually over a limited range of values of the random variable. We find that the PDF evolves as a function of scale $r$, and demonstrate that this evolution is qualitatively different in the dissipative and inertial ranges.

\section{2. Experiment}

The experiments were carried out in a tank $1m \times 1m$ in lateral dimensions. It was filled with water to a depth of $30 cm$. Turbulence was generated by a pump ($8 hp$) which re-circulates water through a system of 36 rotating jets placed horizontally across the tank floor. This system ensured that the source of turbulent injection was far removed from the free-surface where the measurements were made \cite{CDGS2004}. More
importantly, it minimized the amplitude of surface waves which were unavoidable, but have been proven to play no role in the effects reported here \cite{CDGS2004}. The hydrophillic particles chosen here are subject to capillary forces which are very small compared to forces coming from the turbulence, and do not affect the results as they do in \cite{PFL2006, FWDL2005}. The non-inertial character of the particles is minimal because the product $\tau_s \lambda_1$, where $\lambda_1$ is the largest Lyapunov exponent and $\tau_s$ is the stopping time of the particle \cite{BCG2008}, is of the order $5 \times 10^{-4}$, which is too small to lead to any significant inertial effect \cite{BLG2004}.

During an experimental run, particles ($ 50 \mu m$ diameter and specific gravity of 0.25) were constantly seeded into the fluid from the tank floor. From there they underwent turbulent mixing as they rose due to buoyancy and were uniformly dispersed by the time they rose to the surface. Once at the free-surface,
their motion was constrained to the two-dimensional surface plane and they were tracked via a high-speed camera (Phantom v.5) situated above the tank. Instantaneous velocity fields were measured using an in-house developed particle imaging velocimetry (PIV) program which processed the recorded images of the floaters. The scheme of particle injection ensured that both the sources and sinks of the flow received equal
coverage of particles to perform the PIV analysis. The injection scheme was also necessary to replace particles that left the camera field of view, allowing the experiment to be performed for an adequate time period. The experimentally measured velocity vectors were spaced (on average) by $\delta x =$ 2.5 $\eta$. The level of turbulence achieved with our setup leads to a Taylor Microscale Reynolds number $Re_\lambda$ in the range $150-170$. In the flow, the integral length scale is $l_0 \approx 1.45 cm$, the Taylor scale is $\lambda \approx 0.45 cm$ and the dissipative scale is $\eta = (\nu^3/\epsilon)^{1/4} \approx 0.02 cm$, where $\epsilon$ is the energy dissipation rate \cite{frisch}. The camera field-of-view is a square area of side length $L = 9 cm$. The camera's height above the water surface was chosen so that a pixel size is roughly $0.1 mm$, comparable to the dissipative scale of the turbulence.

The measured velocity field was then used to solve the equation of motion for Lagrangian particles : $\frac{d{\bf x}_i}{dt} = {\bf v}({\bf x}_i(t),t)$, where ${\bf v}({\bf x}_i,t)$ is the velocity field and ${\bf x_i}=(x_i,y_i)$ are the individual particle positions. To achieve sub-grid resolution of the measured velocity field, as is required to measure $n_r$ in the dissipative range, the bi-cubic interpolation scheme discussed in \cite{Pope88} was implemented.  To use this scheme it is necessary for the measured velocity grid spacing to satisfy the criterion $\delta x < \pi \eta$, where $\delta x$ is the (above quoted) average measured velocity grid spacing. We have tested to ensure that the results presented here do not depend on the velocity grid spacing by varying the spacing from $\delta x$ = 2.5 $\eta$ to $\delta x$ = 4 $\eta$. All of the statistics presented in this work were obtained by evolving a total number of $N_t \sim 4 \times 10^5$ Lagrangian particles per frame for each experimental run. For each data set, a uniform distribution of Lagrangian tracers was generated at time $ t = 0 s$ and evolved using the above described scheme. The statistics presented here are discussed in terms of length $s$ compared to the dissipative length scale $\eta$, $r=s/\eta$, making $r$ dimensionless.

Concentration statistics in turbulent studies can be studied by dividing the field of view into boxes of size $r$ and counting in each cell the number $N(r,t)$ of particles, thus computing the the value of the corresponding Eulerian (coarse-grained) concentration, $n_r(t) = \frac{N(r,t)}{\langle N(r,t) \rangle}$. Alternatively, the concentration can be measured in the Lagrangian frame. To do this, one draws a box (or circle), of size $r$ around each (moving) particle and counts the number of particles $N^*(r,t)$ that lie inside, including the centered particle. The Lagrangian (dimensionless) concentration is then defined as: $c_r(t) = \frac{N^*(r,t)}{ \langle N(r,t) \rangle}$, where the denominator is the Eulerian frame mean. Physically, averaging a quantity over a set of Lagrangian particles is equivalent to averaging over space, with a weight proportional to the particle density. The relation between the moments of $n_r$ and $c_r$ :

\begin{equation}
\langle c_r^p \rangle = \langle n_r^{p+1} \rangle \label{Eul_Lag}.
\end{equation}
follows from this simple fact, at least when the value of $r$ is small enough, so the correlation functions of the concentration field simply scale with $r$ \cite{BFF99}. Likewise the PDF's of $n_r$ and $c_r$ are related:
$n_r \Pi(n_r) = \Pi(c_r)$. Experimentally, one obtains much more accurate results by using the Lagrangian field $c_r$, rather than $n_r$ \cite{GP83}. We have checked that the accurate results determined with $c_r$ are consistent with the results obtained with $n_r$, using Eq.\ref{Eul_Lag}. In the following, we express our results in terms of the Lagrangian concentration field, $c_r$. The statistics of the concentration field discussed here
have been taken only in the statistically steady-state regime, by systematically leaving out the initial transients, necessary for the moments of the distributions to reach their steady-state.

\begin{table}
\caption{\label{table}Turbulent parameters measured at the surface.  Measurements are made at several values of the $Re_\lambda$ with an average $Re_\lambda \simeq 160$.  The parameters listed are averages, with deviations less than $10\%$.}
\begin{center}
\begin{tabular}{|p{1.5in}|p{1.2in}|p{0.7in}|}
 \hline
 \small {Taylor microscale} & $\lambda $=$\sqrt{\frac{v^{2} _{rms} }{\left\langle \left({\partial v_{x}
\mathord{\left/{\vphantom{\partial v_{x}  \partial x}}\right.\kern-\nulldelimiterspace} \partial x} \right)^{2}
\right\rangle } }$ & 0.47 (cm) \\ \hline
\small{Taylor Re$_\lambda$} & Re$_\lambda$=$\frac{v_{rms} \lambda }{\nu }$ & 160 \\ \hline
\small{Integral Scale} & $l_0$=$\int dr\frac{\left\langle v_{\left\| \right. }
(x+r)v_{\left\| \right. } (x)\right\rangle }{\left\langle
\left(v_{\left\| \right. } (x)\right)^{2} \right\rangle }$ & 1.42 (cm) \\ \hline
    \small{Large Eddy Turnover Time} & $\tau _{0} =\frac{l_{0} }{v_{rms} } $ & 0.43 (s) \\ \hline
    \small{Dissipation Rate} & $\varepsilon_{diss}$=$10\nu \left\langle \left(\frac{\partial v_{x} }{\partial x} \right)^{2} \right\rangle $ & 6.05 (cm2/s3) \\ \hline
    \small{Kolmogorov scale} & $\eta =\left(\frac{\nu ^{3} }{\varepsilon }\right)^{1/4} $ & 0.02 (cm) \\ \hline
    \small{RMS Velocity} & $v_{rms} =\sqrt{\left\langle v^{2} \right\rangle-\left\langle v\right\rangle ^{2} }$ & 3.3 (cm/s) \\ \hline
    \small{Compressibility} & ${\cal C}=\frac{\left\langle \left(\stackrel{\rightharpoonup}{\nabla }_{2} \cdot \stackrel{\rightharpoonup}{v}\right)^{2} \right\rangle }{\left\langle \left(\stackrel{\rightharpoonup}{\nabla }_{2} \stackrel{\rightharpoonup}{v}\right)^{2} \right\rangle }$ & 0.49 $\pm$ 0.02\\
    \hline
  \end{tabular}
\end{center}
\end{table}

\section{3. Results}

\begin{figure}
\begin{center}
\includegraphics[width = 3.5 in]{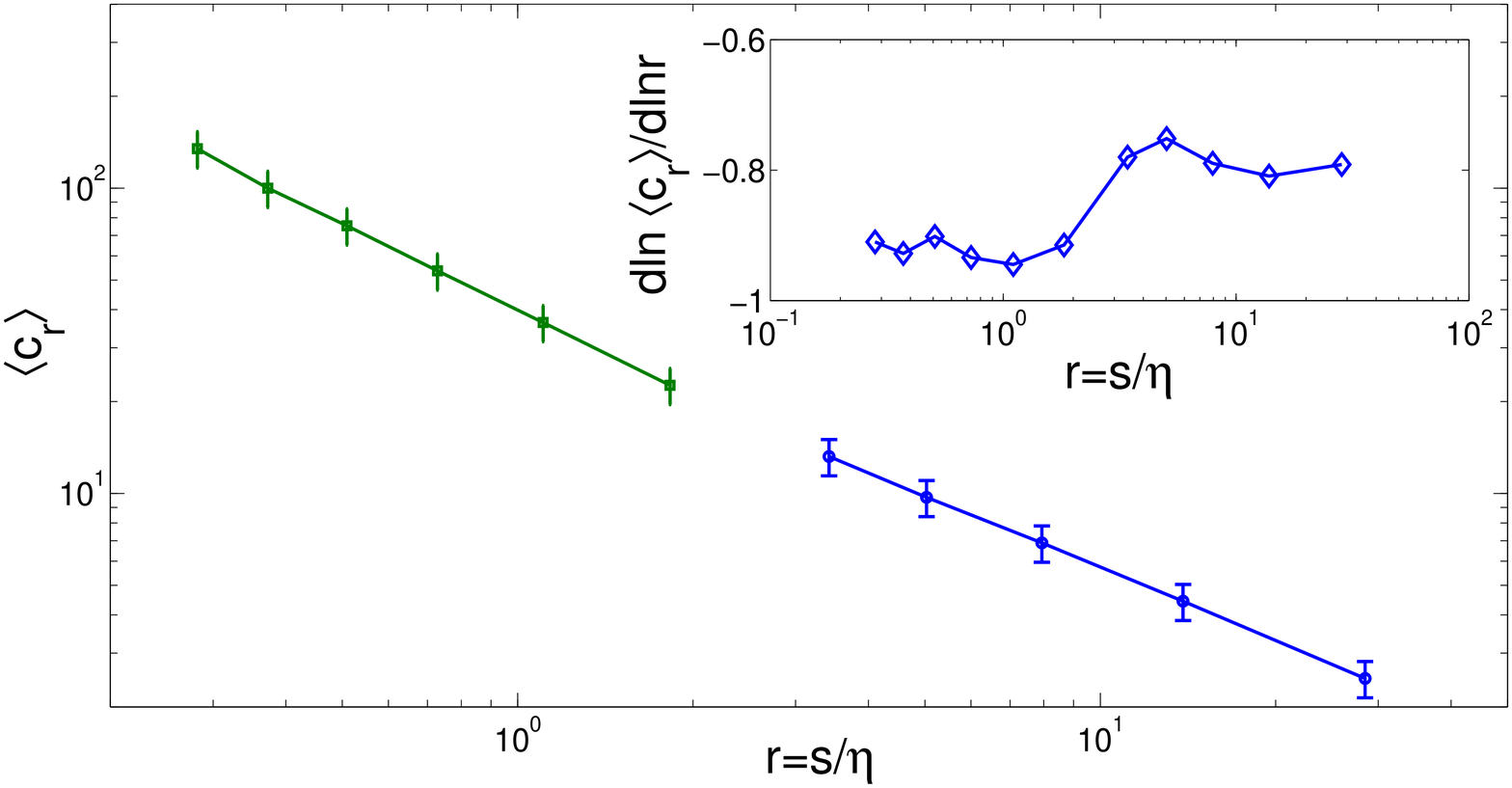}
\end{center}
\caption{(Color online) First moment of the Lagrangian concentration field $c_r$. The average $\langle c_r \rangle$ exhibits two scaling regimes. In the dissipative range, for $s \lesssim 3 \eta$, $ \langle c_r \rangle $ scales as $r^{-\alpha_d}$ with $\alpha_d = 0.92\pm0.02$, whereas for $s \gtrsim 5 \eta$, it scales as $r^{-\alpha_i}$ with $\alpha_i = 0.79\pm0.03$. The inset shows the derivative $\frac{d \ln \langle c_r \rangle}{d r }$, and demonstrates the quality of the reported power-law behavior. }
\label{cr_1}
\end{figure}

We discuss the properties of the steady-state distribution of the coarse-grained particle concentration $c_r$ defined above. We begin with the lowest non trivial moment : $\langle c_r \rangle$, and its dependence on the dimensionless coarse-graining scale $r$ \cite{BCG2008}. Scaling relations, such as $\langle c_r \rangle \sim r^{-\alpha}$, provides a characterization of the topological structure of the particle distribution \cite{GP83}.
For simple geometrical structures such as points, lines, and surfaces, the scaling exponent $\alpha$ take on the simple integer values $\alpha = 2,1,0$, respectively. For more complicated structures, as in our case, $\alpha$ is a non-integer and gives a quantitative measure of the complex topological structure. Fig. \ref{cr_1} shows
that $\langle c_r \rangle$ exhibits two different power-law behaviors, at small ($r \lesssim 3$) and at larger ($r \gtrsim 5$) scales. These data can be fit with an exponent $\alpha_d = 0.92\pm0.02$ ($\alpha_i = 0.79\pm0.03$) in the inertial (dissipative) ranges. According to Fig. \ref{cr_1}, the dissipation range extends up to scales larger than $\eta$ (for $r \approx  3-5$). This is consistent with the fact that the dissipative range crossover in many experiments occurs at $r > 1$ \cite{Meneveau96,Benzi93}. The difference between the two scaling domains can be best seen by plotting the derivative $d \ln \langle c_r \rangle / d \ln r$, see the inset in Fig. \ref{cr_1}. The qualitative impression, see Fig.\ref{particles}, that the net result of the coagulation process are line-like structures is completely consistent with the values of the exponents close to $1$. Higher order moments of the coarse-grained concentration also exhibit unique scaling regimes in the inertial and d
 issipative ranges, transitioning between the two regimes at approximately $5 \eta$. Our results are thus in qualitative agreement with the observations of \cite{DP2008} for $\langle n_r^2 \rangle$, which also show two separate scaling intervals corresponding to the inertial and dissipative ranges.

\begin{figure}
\includegraphics[width = 3.5 in]{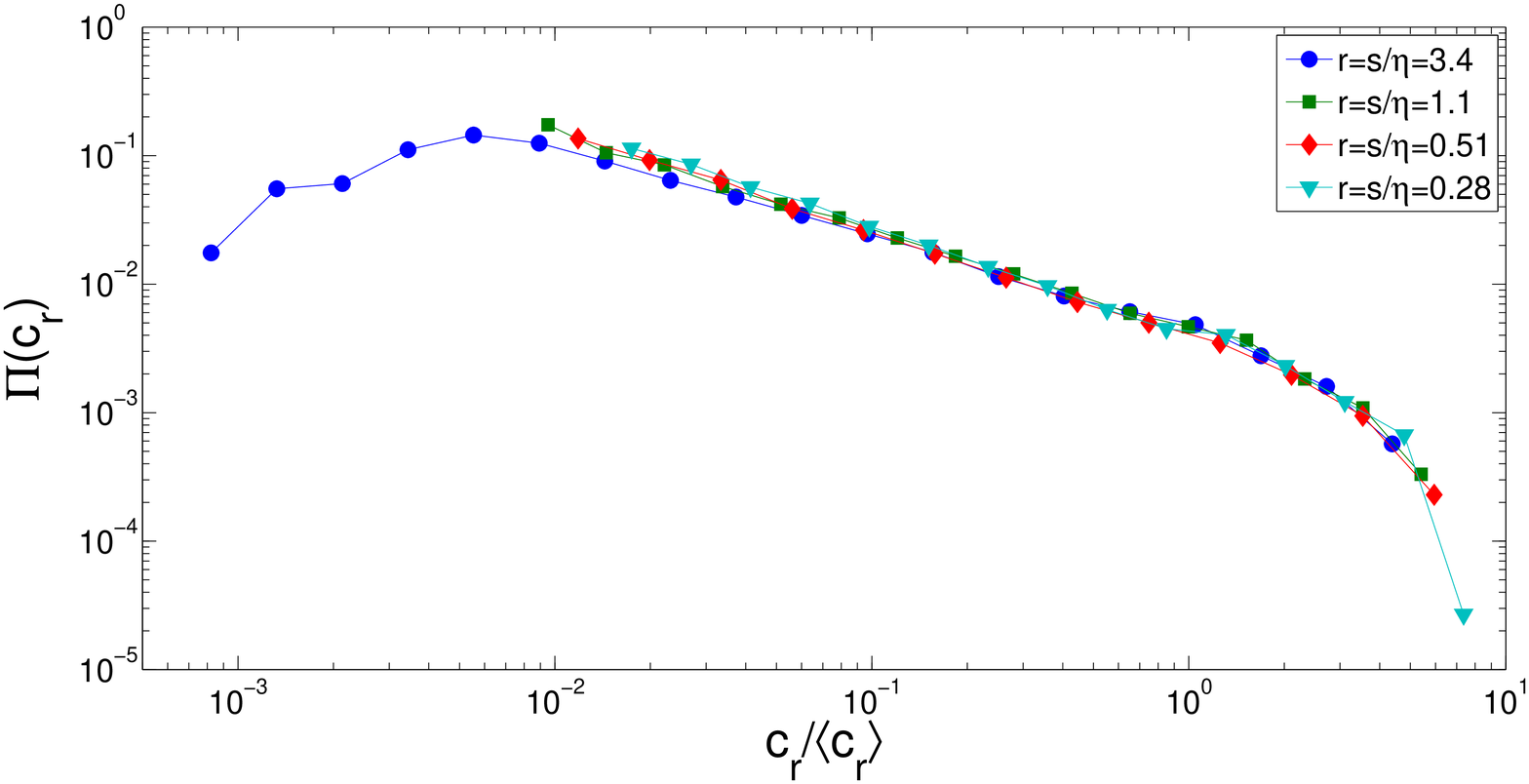}
\includegraphics[width=3.5 in]{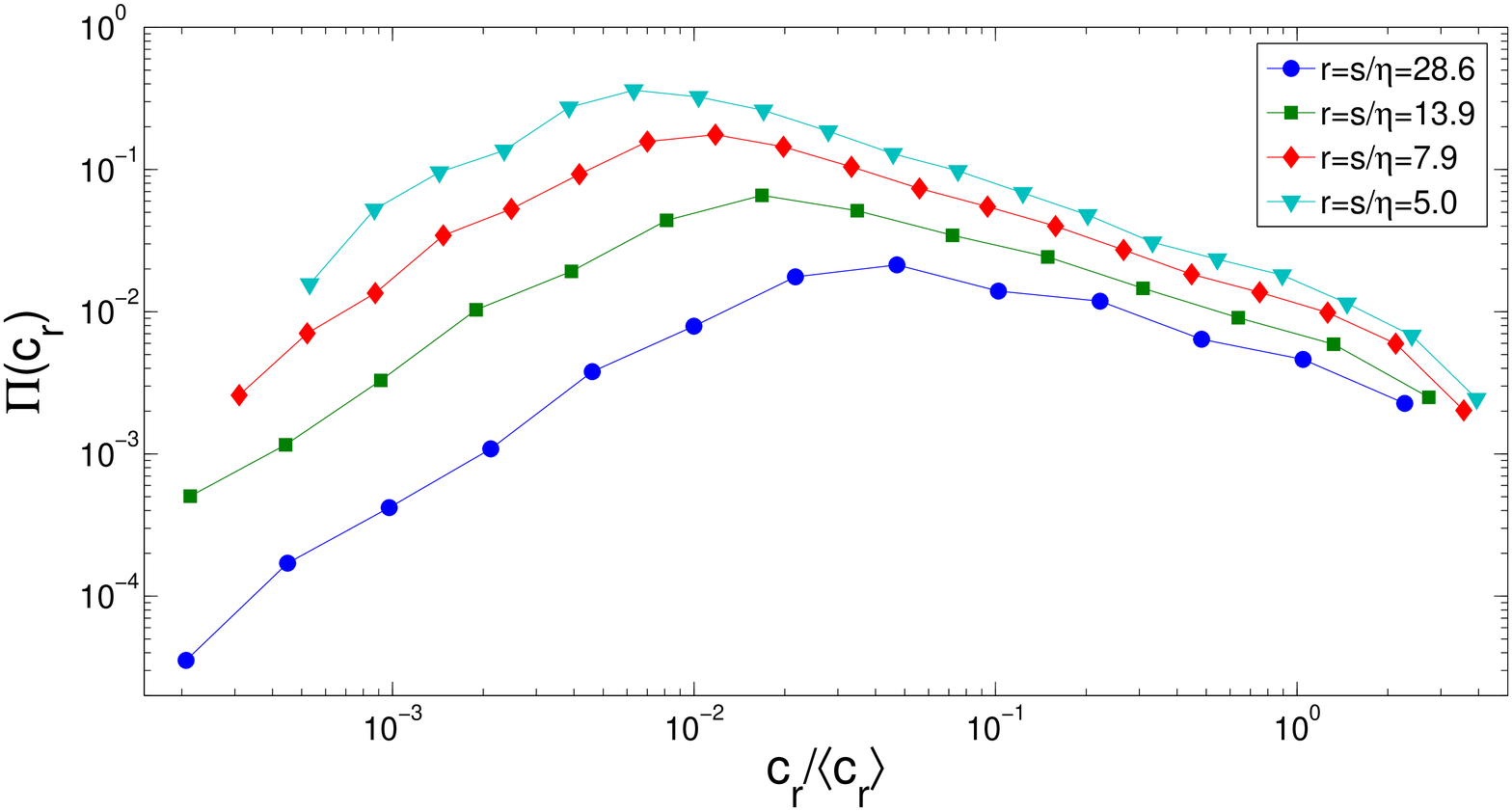}
\caption{(Color online) Top panel : $\Pi ( c_r )$ in the dissipative range $r \lesssim 3$. The PDFs of the concentration, scaled by its mean value $\langle c_r \rangle$ exhibit a power-law distribution at low value of $c_r/\langle c_r \rangle$: $\Pi(x) \propto x^{-\beta}$ with $\beta \approx 0.8 \pm 0.05$. The PDFs decay faster than algebraically for
$c_r/\langle c_r \rangle \gtrsim 1$. Bottom panel : $\Pi ( c_r )$ in the inertial range $r \gtrsim 5$. The PDFs in the inertial range also exhibit a power-law distribution at low value of $c_r/\langle c_r \rangle$ greater than the PDF's maximum : $\Pi(c_r) \propto c_r^{-\beta_r}$ whose exponent $\beta_r$ decreases when $r$ increases. Here, $\beta_r \simeq 0.5 $ at $r=28.6$ and increases to $\beta_r \simeq 0.75 $ at $r=5.0$  The decay of the PDF at large values of $c_r/\langle c_r \rangle \gtrsim 1$ is faster than algebraic, clearly seen for $r=7.9$ and $5.0$. The PDFs in the inertial range have been shifted vertically to better see the evolution of the power-law region.}
\label{PDF}
\end{figure}

We now discuss the PDF $\Pi(c_r)$ of the concentration. Fig. \ref{PDF} shows the PDFs of $c_r$, normalized
by the mean $\langle c_r \rangle$, in the dissipative and inertial ranges. In the inertial range, and for $r=3.4$ in the dissipative range, the PDFs display a clear maximum in the distribution, with this maximum occuring at smaller values of $c_r/\langle c_r \rangle$ as $r$ decreases. For values of $c_r/\langle c_r \rangle$ less than this maximum, the PDFs exhibit a plausibly linear behavior : $\Pi(c_r) \propto c_r$, which corresponds to a PDF in the Eulerian frame $\Pi(n_r) \propto constant$ (see Eq. (\ref{Eul_Lag})). This ensures that the PDF is normalizable in both the Lagrangian and Eulerian frames. This very small $c_r/\langle c_r \rangle$ wing of the PDF is not observed deep in the dissipative range, for $ r \le 1.1$, due to the fact that values of $c_r/\langle c_r \rangle$ corresponding to less than one particle are not adequately resolved.

The most interesting aspect of the PDFs in the dissipative (inertial) range is the power-law behavior at low (intermediate) values of $c_r/\langle c_r \rangle$ greater than the PDF's maximum. This algebraic decay corresponds to the depleted regions seen in Fig. \ref{particles} rather than the string-like coagulations. In both spatial ranges, the PDFs can be characterized by a power-law behavior, $\Pi(c_r) \propto c_r^{-\beta_r}$, over a limited range of $c_r/\langle c_r \rangle$ followed by a faster than algebraic fall-off at large $c_r/\langle c_r \rangle$. The variation of the exponent $\beta_r$ in the dissipative range, Fig.\ref{PDF}, is very small : $\beta_r \approx 0.8 \pm 0.05$ for $r \le 3.4$. In comparison, the exponent $\beta_r$ decreases when the scale $r$ increases in the inertial range (see Fig. \ref{PDF}), with $\beta_r \simeq 0.5 $ at $r=28.6$ and $\beta_r \simeq 0.75$ at $r=5.0$. The fast decay of the PDFs at large $c_r/\langle c_r \rangle$ ensures that all the
  moments of the distribution $\langle c_r^p \rangle$ exist, for all $p \ge 1$.

The PDFs in both the inertial and dissipative ranges show a systematic variation of their form as a function of scale $r$, qualitatively similar to the evolution of the velocity increment PDF in turbulence \cite{frisch}. In the dissipative range, this observation is consistent with the theoretical arguments of \cite{BGH2004}, leading to
the prediction that the particle distribution is multi-fractal. Such a multi-fractal property implies that the PDF must depend on the scale $r$. The change in the PDF with scale $r$ mostly affects the large $c_r/\langle c_r \rangle$ behavior, where the PDF decays faster than algebraically. The pre-factor of the algebraic behavior of the PDF also varies with scale $r$.

\section{4. Summary}
To summarize, we have characterized the statistical properties of particle clustering on the surface of a turbulent flow by measuring the coarse-grained concentration around {\it each} particle in the system. Our study allows us to cover a range of scales extending from dissipative up to inertial scales. The first moment of the particle distribution $\langle c_r \rangle$ exhibits two different scaling regimes in the dissipative and in the inertial ranges, for scales smaller and larger than $\sim 5 \eta$. The PDF of $c_r$ vary systematically with scale. Indeed, our limited results are consistent with a multi-fractal scaling both in the dissipative and in the inertial ranges, which is generally consistent with the findings of \cite{DP2008}. The PDFs exhibit power-law behavior at small values of $c_r/\langle c_r \rangle$, characterizing the very high probability of having regions containing very low concentrations.

The corresponding distribution of the Eulerian concentration $n_r$ behaves as $\Pi(n_r) \propto n_r^{-(1 + \beta_r)}$, which is not even normalizable when $n_r \rightarrow 0$, thus implying saturation at a small value of $n_r$, as indeed suggested by Fig. \ref{PDF}. Such a behavior has been observed in \cite{PFL2006} with $\beta_r = 1$. The explanation in terms of caustics \cite{WM2005} proposed in \cite{PFL2006} is unlikely to apply here, both because our particles strictly follow the Lagrangian evolution equation, and because $\beta_r \ne 1$.

Our work thus indicates that the intermittent distribution of particles in free-surface flows is characterized both by very high particle concentration along string-like regions as well as by a strong depletion of particles in other parts. A proper description of the resulting power-law tails requires a better understanding of the very efficient
expulsion mechanism of particles from large regions of the flow. Understanding this particle clustering is of fundamental interest and practical use in a range of surface transport phenomena, such as the dispersion of phytoplankton or contaminants in oceanic flow  \cite{R1926,S1949,A2000}.

We acknowledge very helpful discussions with G. Falkovich and K. Gawedzki. Funding was provided by the US National Science Foundation grant No. DMR-0604477. and by the French ANR (contract DSPET), and by IDRIS. This work was partially carried out under the auspices of the National Nuclear Security Administration of the U.S. Department of Energy at Los Alamos National Laboratory under Contract No. DE-AC52-06NA25396.

\bibliography{conc30oct09_JL}
\end{document}